\documentclass[reprint, prl, aps,
amsmath,amssymb,showpacs,superscriptaddress]{revtex4-1}
\pdfoutput=1

\usepackage{siunitx}
\usepackage[breaklinks=false]{hyperref}
\usepackage{graphicx}
\usepackage{braket}
\usepackage{verbatim}
\usepackage{color}
\usepackage[normalem]{ulem}

\begin{document}
\title{The AC Stark Effect in Ultracold Polar $^{87}$Rb$^{133}$Cs Molecules}

\author{Philip D. Gregory}
\affiliation{Joint Quantum Centre (JQC) Durham-Newcastle, Department of
Physics, Durham University, South Road, Durham DH1 3LE, United Kingdom}

\author{Jacob A. Blackmore}
\affiliation{Joint Quantum Centre (JQC) Durham-Newcastle, Department of
Physics, Durham University, South Road, Durham DH1 3LE, United Kingdom}

\author{Jesus Aldegunde}
\affiliation{Departamento de Quimica Fisica, Universidad de Salamanca, 37008
Salamanca, Spain}

\author{Jeremy M. Hutson}
\email{J.M.Hutson@durham.ac.uk} \affiliation{Joint Quantum Centre (JQC)
Durham-Newcastle, Department of Chemistry, Durham University, South Road,
Durham, DH1 3LE, United Kingdom}

\author{Simon L. Cornish}
\email{S.L.Cornish@durham.ac.uk} \affiliation{Joint Quantum Centre (JQC)
Durham-Newcastle, Department of Physics, Durham University, South Road, Durham
DH1 3LE, United Kingdom}

\begin{abstract}
We investigate the effect of far-off-resonant trapping light on ultracold
bosonic $^{87}$Rb$^{133}$Cs molecules. We use kHz-precision microwave
spectroscopy to measure the differential AC~Stark shifts between the ground and
first excited rotational levels of the molecule with hyperfine-state
resolution. We demonstrate through both experiment and theory that coupling
between neighboring hyperfine states manifests in rich structure with many
avoided crossings. This coupling may be tuned by rotating the polarization of
the linearly polarized trapping light. A combination of spectroscopic and
parametric heating measurements allows complete characterization of the
molecular polarizability at a wavelength of 1550~nm in both the ground and first
excited rotational states.
\end{abstract}

\date{\today}

\maketitle
Ultracold molecules in optical traps and optical lattices have many potential
applications, ranging from quantum-state-controlled chemistry~\cite{Krems:2008,
Ospelkaus:2010b, Ni:2010, Miranda:2011}, to quantum
simulation~\cite{Santos:2000, Baranov:2012} and quantum
information~\cite{DeMille:2002, Yelin:2006}. Many of these applications rely on
coherent microwave transfer between rotational states of the molecules.
However, all the molecules that have been prepared at ultracold temperatures so
far~\cite{Ni:2008, Takekoshi:2014, Molony:2014, Park:2015, Guo:2016} have
nuclei with non-zero spins, resulting in complex hyperfine and Zeeman
structure~\cite{Aldegunde:2008, Aldegunde:spectra:2009, Aldegunde:nonpolar:2009, Ospelkaus:2010, Will:2016, Gregory:2016}. In such cases, the
laser fields used to confine the molecules have important effects through the
AC Stark effect, particularly for molecules in rotationally excited states. A
thorough understanding of these effects is essential in order to eliminate
differential Stark shifts detrimental to internal state transfer and thus to
develop ultracold polar molecules into a controllable resource for use in
quantum science.

The key quantity that determines the AC Stark effect is the molecular
polarizability. Following a theoretical proposal by Kotochigova and
DeMille~\cite{Kotochigova:2010}, Neyenhuis \emph{et al.}\ \cite{Neyenhuis:2012}
carried out parametric heating experiments at fixed laser intensity to
determine the polarizabilities of different molecular states. They showed that
there exists a \emph{magic} angle for the linear polarization of optical
trapping light. In analogy with magic-wavelength traps in atomic systems, at
the magic angle, the AC Stark shift of hyperfine levels of different rotational
states are the same. This allowed Ramsey interferometry between two rotational
levels of the molecule with a reasonably long coherence time, and led to a
pivotal study of the dipolar spin-coherence time in a 3D optical
lattice~\cite{Yan:2013}. More recently, Dei\ss~\emph{et al.} \cite{Deib:2014,
Deib:2015} have carried out parametric heating experiments on aligned triplet
Rb$_2$ molecules and extracted both isotropic and anisotropic polarizabilities.

In this letter, we explore the dependence of the AC Stark effect on laser
intensity both experimentally and theoretically, using microwave spectroscopy
of the chemically stable and bosonic $^{87}$Rb$^{133}$Cs molecule. We show that
there is a subtle interplay between the AC Stark effect and the hyperfine
structure. The trapping light couples neighboring hyperfine states, giving rich
and complex structure with many avoided crossings as a function of laser
intensity. We use our measurements to extract a precise value for the
anisotropic component of the molecular polarizability. We complete the
characterization of the polarizability tensor by performing parametric heating
and spectroscopic measurements to extract the isotropic component of the
polarizability. Our findings will allow us to engineer trapping potentials
suitable for internal state transfer with long coherence times and have
implications for experiments with other molecular species.

The AC Stark effect arises from the interaction of the electric field due to a
laser of intensity $I$ with the polarizability $\alpha$ of an atom or molecule,
and results in a perturbation in energy of $-\alpha I$. By contrast with the
atomic case, the molecular polarizability is anisotropic; in the case of a
linear diatomic molecule like $^{87}$Rb$^{133}$Cs, the highest polarizability
is along the internuclear axis. We apply a magnetic field $B_{z}$ in the
vertical $z$ direction, and the orientation of the molecule is defined with
respect to this magnetic field. The polarizability of the molecule at an angle
$\theta$ to the internuclear axis is
\begin{equation}\label{eq:Polarizability}
\begin{split}
\alpha(\theta) &= \alpha_{\parallel}\cos^{2}\theta + \alpha_{\perp}\sin^{2}\theta \\
							 &= \alpha^{(0)} + \alpha^{(2)}P_{2}(\cos\theta),
\end{split}
\end{equation}
where $\alpha_{\parallel}$ and $\alpha_{\perp}$ are the polarizability parallel
and perpendicular to the internuclear axis respectively,
$\alpha^{(0)}=\frac{1}{3}(\alpha_{\parallel}+2\alpha_{\perp})$ and
$\alpha^{(2)}=\frac{2}{3}(\alpha_{\parallel}-\alpha_{\perp})$. The trapping
light is linearly polarized in the $xz$ plane at an angle $\beta$ to the
magnetic field. We therefore rotate the polarizability tensor through an angle
$\beta$ and find matrix elements
\begin{equation} \label{eq:MatrixElements}
\begin{split}
& \braket{N',M_{N}'|I\alpha|N,M_{N}} = I\alpha^{(0)} \delta_{NN'} \delta_{M_N M'_N}\\
+& I\alpha^{(2)}\sum_{M}{d_{M0}^{2}(\beta)} (-1)^{M'_{N}}\sqrt{(2N+1)(2N'+1)} \\
&
\begin{pmatrix}
N' & 2 & N \\
0& 0 & 0 \\
\end{pmatrix}
\begin{pmatrix}
N' & 2 & N \\
-M'_{N}& M & M_{N} \\
\end{pmatrix},
\end{split}
\end{equation}
where $N$ is the rotational angular momentum of the molecule, with projection
$M_{N}$ along the magnetic field axis, and $d^{2}(\beta)$ is a reduced rotation
matrix.

To calculate the hyperfine levels in the presence of an AC Stark effect, we
construct the Hamiltonian matrix in a decoupled basis $|NM_N\rangle |I_{\rm
Rb}m_{I}^{\text{Rb}}\rangle |I_{\rm Cs}m_{I}^{\text{Cs}}\rangle$, where $I_{\rm
Rb}=3/2$, $I_{\rm Cs}=7/2$ and $m_{I}^{\text{Rb}}$, $m_{I}^{\text{Cs}}$ are the
corresponding projections. We supplement the Zeeman and hyperfine matrix
elements of ref.\ \cite{Aldegunde:2008} with the AC Stark terms of Eq.\
\ref{eq:MatrixElements}, which are diagonal in and independent of
$m_{I}^{\text{Rb}}$ and $m_{I}^{\text{Cs}}$. We include all basis functions
with $N\le3$ in the calculation. Diagonalizing the resulting Hamiltonian gives
us both energy levels and wavefunctions in the presence of off-resonant
trapping light. We then use the wavefunctions to calculate spectroscopic
transition strengths for the required polarization of microwaves.

The representation of $\alpha{\theta}$ in terms of $\alpha^{(0)}$ and $\alpha^{(2)}$ conveniently
separates the effects of the two components of the polarizability. The
isotropic component $\alpha^{(0)}$ shifts all diagonal matrix elements by the
same amount and has no effect on transition frequencies, though it does
contribute to optical trapping. The anisotropic component $\alpha^{(2)}$, on
the other hand, mixes different hyperfine states through matrix elements
diagonal and off-diagonal in $M_N$ and dependent on $\beta$. For $N=0$, the
matrix elements of $\alpha^{(2)}$ are zero, so the polarizability is simply
$\alpha^{(0)}$ for all hyperfine states. For $N=1$, however, $\alpha^{(2)}$ has
important effects; if we neglect terms off-diagonal in $N$, the matrix of the
polarizability tensor between
basis functions with $N=1$, $M_N=0$, +1 and $-1$ is
\begin{equation} \label{eqn:PolMatrix}
\begin{split}
&\braket{1,M_{N}'|I\alpha|1,M_{N}} = I\alpha^{(0)} + \\
& \frac{I\alpha^{(2)}}{5}
\begin{pmatrix}
2P_{2}(\cos\beta)							& -\frac{3}{\sqrt{2}}\sin\beta\cos\beta		& +\frac{3}{\sqrt{2}}\sin\beta\cos\beta \\
-\frac{3}{\sqrt{2}}\sin\beta\cos\beta		& -P_{2}(\cos\beta) 						& -\frac{3}{2}\sin^{2}\beta \\
+\frac{3}{\sqrt{2}}\sin\beta\cos\beta		& -\frac{3}{2}\sin^{2}\beta  				& -P_{2}(\cos\beta)  \\
\end{pmatrix}.
\end{split}
\end{equation}


In the absence of the trapping laser,
$M_F=M_N+m_{I}^{\text{Rb}}+m_{I}^{\text{Cs}}$ is a good quantum number, but
$M_N$, $m_{I}^{\text{Rb}}$ and $m_{I}^{\text{Cs}}$ are not individually
conserved. When the trap laser is polarized along $B_z$, corresponding to
$\beta=0$, $M_F$ is still conserved. However, when $\beta\ne0$, the AC Stark
effect mixes levels with different values of $M_F$ and there are no good
quantum numbers except reflection symmetry in the $xz$ plane.

Our experimental apparatus and method for creating ultracold
$^{87}$Rb$^{133}$Cs molecules have been discussed in previous
publications~\cite{Molony:2014, Harris:2008, Jenkin:2011, Cho:2011,
McCarron:2011, Koppinger:2014, Gregory:2015, Molony:2016}. In this work we
create a sample of up to $\sim3000$ $^{87}$Rb$^{133}$Cs molecules in their
absolute ground state at a temperature of $\sim1~\mu$K by magnetoassociation on
a Feshbach resonance~\cite{Koppinger:2014} followed by transfer to the
hyperfine and rovibronic ground state by stimulated Raman adiabatic passage
(STIRAP)~\cite{Molony:2014}. The transfer of the molecules to the ground
state is performed in free space~\cite{Molony:2016}. We recently reported the
coherent control of the rotational and hyperfine state of the molecules using
external microwave fields, also in free space~\cite{Gregory:2016}.

\begin{figure}[tb]
\centering
\includegraphics[width=\linewidth]{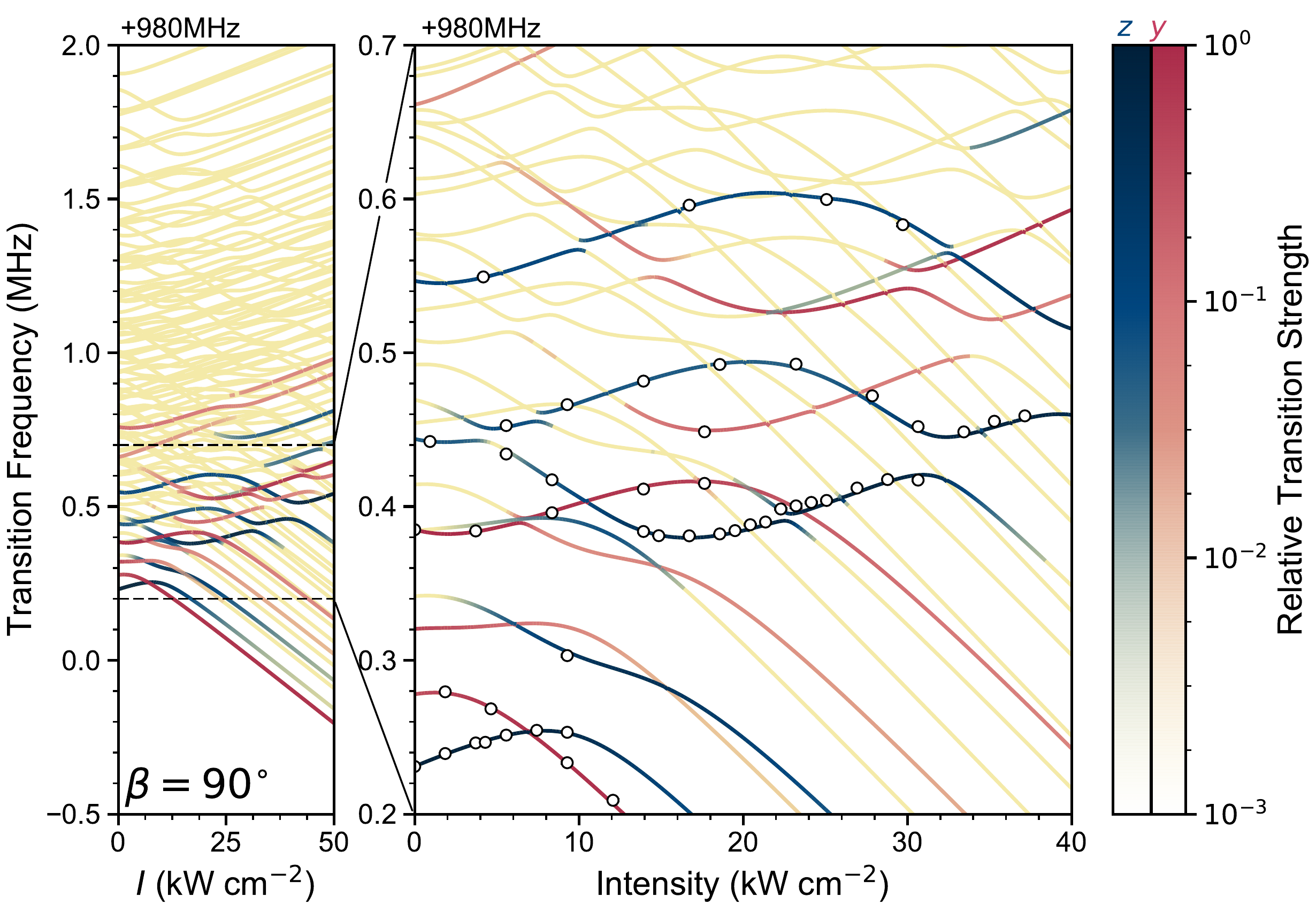}
\caption{\label{fig:Beta90} Transition frequencies from the lowest-energy
hyperfine state of the rovibronic ground state $N=0, M_{F}=5$ to the 96
hyperfine states for $N=1$ as a function of laser intensity, for laser
polarization perpendicular to the magnetic field, $\beta=90^{\circ}$. The
relative transition strengths for microwaves polarized along $z$ and $y$ and
are shown as blue and red color maps respectively. The data points show
experimental microwave frequencies.}
\end{figure}

To measure the differential AC~Stark shift between $N=0$ and $N=1$, we perform
microwave spectroscopy in the presence of the dipole trapping light. A single
beam ($\lambda=1550$~nm, waist = 95~$\mu$m) is switched on for 500~$\mu$s
before the microwave pulse to allow the intensity of the trapping light to
stabilize. The laser polarization is tunable to a precision of $\pm1^{\circ}$
by a $\lambda/2$ waveplate. The molecules experience a reasonably homogeneous
laser intensity within 2\% of the peak intensity. With the molecules initially
in the spin-stretched rotational and hyperfine ground state ($N=0, M_{N}=0,
m_{I}^{\text{Rb}}=3/2, m_{I}^{\text{Cs}}=7/2)$, we pulse on the microwave field
for a time ($t_{\text{pulse}}$) that is less than the duration of a $\pi$ pulse
for the relevant transition. We measure the number of molecules remaining in
the ground state by reversing the STIRAP sequence to dissociate the molecules
and using absorption imaging to detect the resulting atoms. We observe the
microwave transition as an apparent loss of ground-state molecules. All the features 
we measure are Fourier-transform limited, with widths
proportional to $1/t_{\text{pulse}}$. The microwave power is tuned to allow
pulse times of 100-180~$\mu$s, yielding Lorentzian lines with 5-10~kHz full
width at half maximum. We observe transitions due to microwaves polarized along
the $z$ and $y$ axes.


We begin by setting the laser polarization perpendicular to the magnetic field,
i.e. $\beta=90^{\circ}$. Fig.~\ref{fig:Beta90} shows the measured transition
frequencies for $B_z = 181.5$\,G as a function of laser intensity. These are
superimposed on calculations of the 96 hyperfine levels of $N=1$, using
molecular constants fitted to the experiments as described below. Calculated transition strengths are indicated with blue and red shading for microwaves polarized along $z$ and $y$, respectively. Many avoided crossings appear in the region where $I\alpha^{(2)}$ is
comparable to the hyperfine couplings and Zeeman splittings. The basis
functions that carry the spectroscopic intensity cut through the manifold of
states, resulting in a complicated variation in transition strengths as each
state brightens and fades. At sufficiently high laser intensities, the AC Stark
effect dominates the Zeeman splittings; $N$ eventually requantizes along the
laser polarization axis, and the pattern of transition strengths and
frequencies simplifies.

\begin{figure}[tb]
\centering
\includegraphics[width=\linewidth]{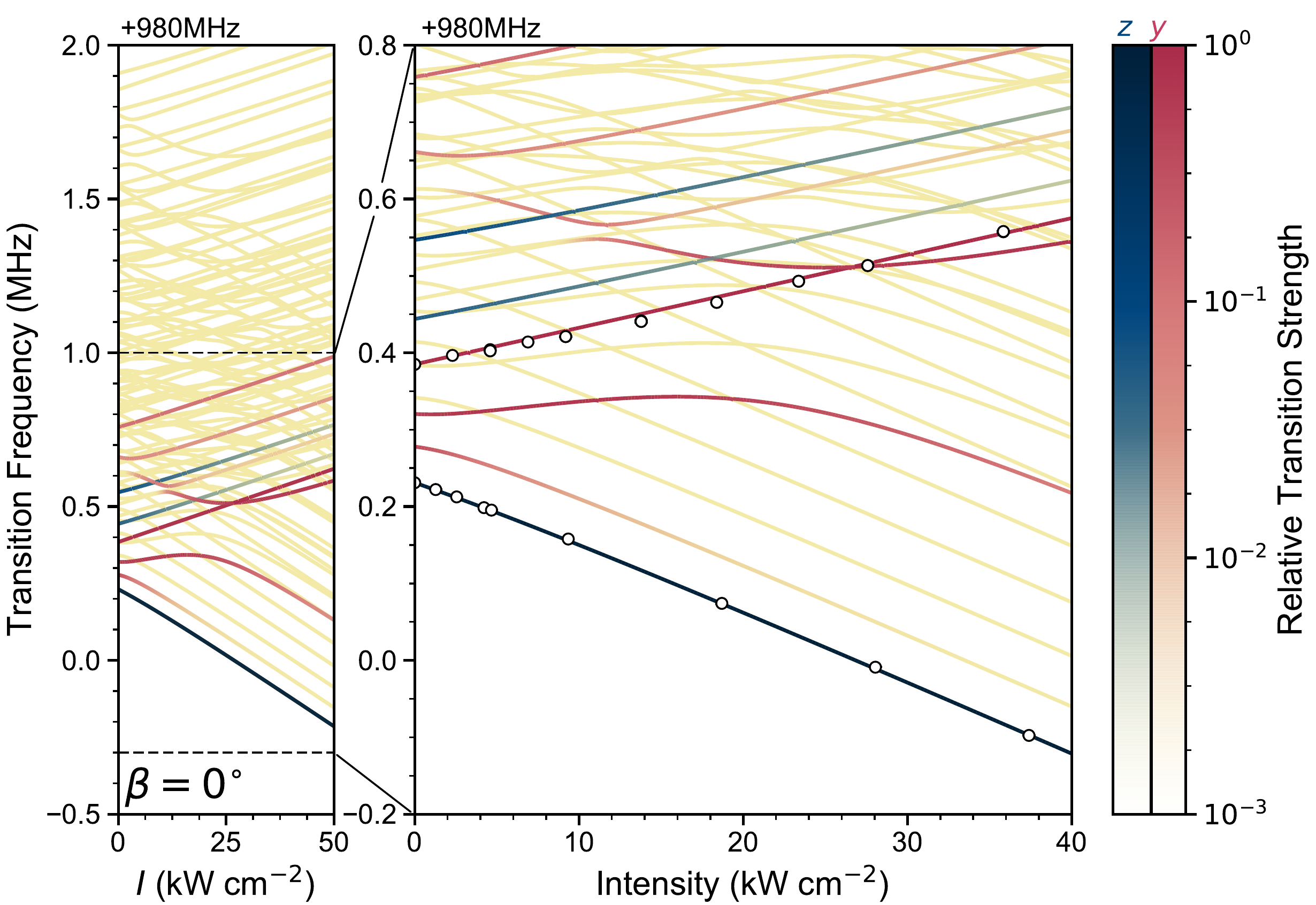}
\caption{\label{fig:Beta00} Transition frequencies from $N=0, M_{F}=5$ to $N=1$
as a function of laser intensity, for laser polarization parallel to the
magnetic field, $\beta=0^{\circ}$. The relative transition strengths are coded
as in Fig.~\ref{fig:Beta90}. The data points show experimental microwave
frequencies.}
\end{figure}


Fig.~\ref{fig:Beta00} shows analogous results for the laser polarization parallel
to the applied magnetic field, $\beta=0^{\circ}$. A single beam is used at low
intensities, as in Fig.~\ref{fig:Beta90}. At the highest intensities shown,
however, two beams are used to form a crossed optical dipole trap. The beams
propagate in the $xy$ plane and cross at an angle of $\sim27$$^\circ$. In this
case $M_F$ is a good quantum number even in the presence of the trapping laser.
The three $M_F=5$ hyperfine states for $N=1$ (blue) diverge as a function of
laser intensity, and there are no avoided crossings between them. At high
intensity $M_N$ becomes an increasingly good quantum number, and the two states
with $M_N=\pm1$ lose intensity for microwaves polarized along $z$.
Nevertheless, strong avoided crossings still exist where states with the same
$M_{F}$ cross.

\begin{figure}[tb]
\centering
\includegraphics[width=\linewidth]{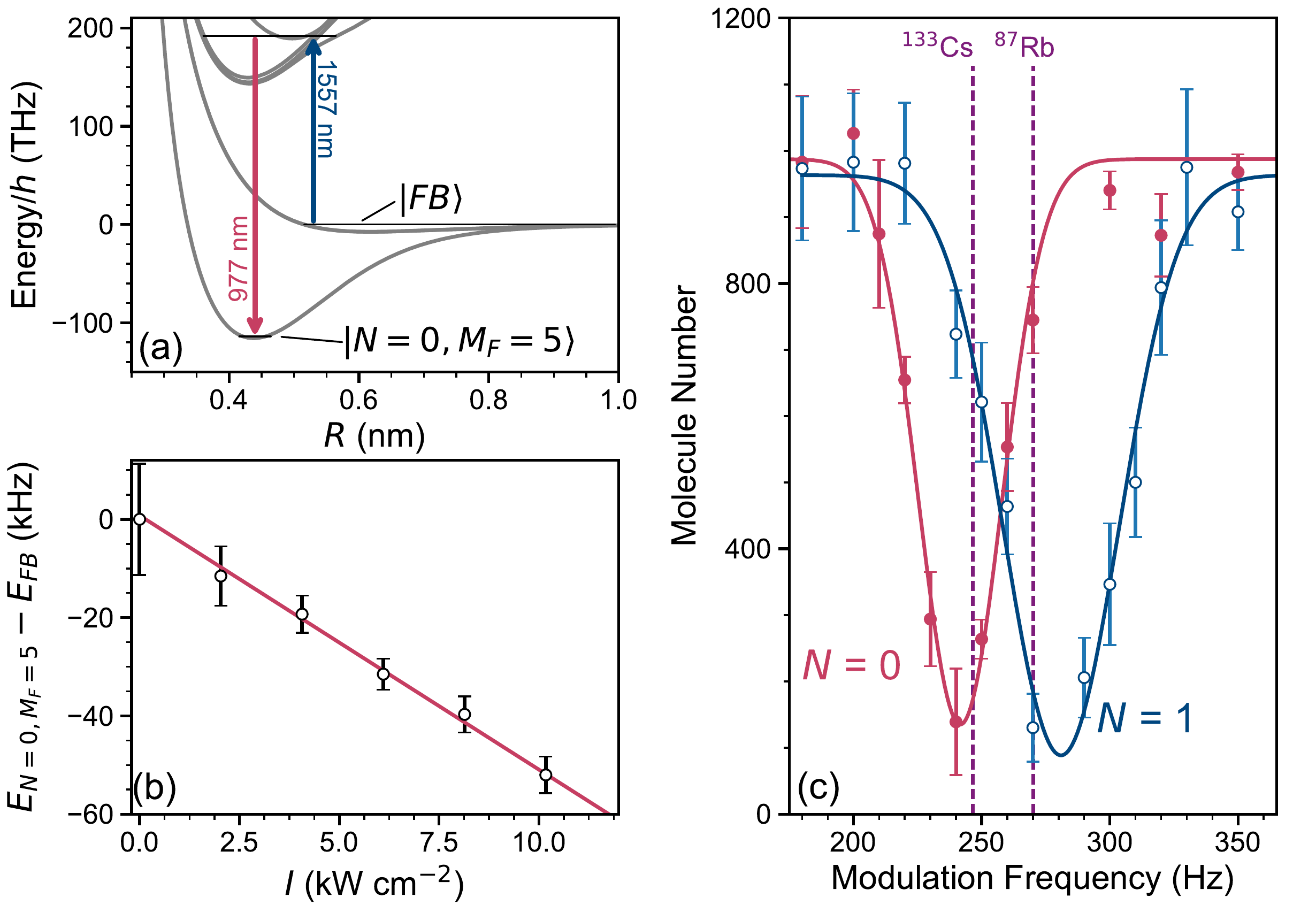}
\caption{\label{fig:Isotropic} Measurement of the isotropic component
$\alpha^{(0)}$. (a)~Transitions used in STIRAP of molecules to the lowest-energy
hyperfine state of the rovibronic ground state. (b) AC Stark shift of the
STIRAP dark-state resonance. (c)~Parametric heating measurements of trap
frequency for molecules in the lowest-energy hyperfine states of $N=0$ (red
closed circles) and $N=1$ (blue open circles) in traps with the same peak laser
intensity. The dotted lines show center frequencies for $^{87}$Rb and
$^{133}$Cs atoms. }
\end{figure}

The experimental uncertainties are not visible on the scale presented in
Figs.~\ref{fig:Beta90} and~\ref{fig:Beta00}. The statistical uncertainties in
the transition frequencies are typically $\pm0.5$~kHz. The dominant uncertainty
in the laser intensity on the other hand is systematic and due to the
uncertainty in the beam waist at the position of the molecules. We estimate
this uncertainty to be $\pm3\%$ of the peak intensity.

We have fitted the experimental results of Figs.~\ref{fig:Beta90}
and~\ref{fig:Beta00} independently to obtain $\alpha^{(2)}$, while holding the
hyperfine constants fixed at the values of ref.\ \cite{Gregory:2016}. For laser
polarizations $\beta=0^{\circ}$ and $90^{\circ}$, we obtain values of
$\alpha^{(2)}_{\beta=0^{\circ}}/4\pi\epsilon_0=507(1)\,a_0^3$ and
$\alpha^{(2)}_{\beta=90^{\circ}}/4\pi\epsilon_0=602(2)\,a_0^3$ respectively.
The uncertainties given are the statistical uncertainties found during fitting,
but both results are subject to the same systematic uncertainty in intensity
described above. Despite the difference between the two fitted values, the
theory in each individual case describes the observed AC~Stark shift well.

To characterize the polarizability of the molecule fully, we must also measure
the \emph{isotropic} component $\alpha^{(0)}$. To do this, we perform several
complementary measurements. First, we observe the energy shift of $N=0$,
$M_F=5$ with respect to an initial weakly bound Feshbach state. This is given
by the AC Stark shift of the two-photon transition used in
STIRAP~\cite{Molony:2016b}, shown in Fig.~\ref{fig:Isotropic}(a) and (b). This
energy shift gives the difference in polarizability between the two molecular
states, and the polarizability of the Feshbach state is simply the sum of the
polarizabilities of the constituent atoms, which are well
known~\cite{Safronova:2006}.

In addition to this spectroscopic method, we perform parametric heating on the
molecular sample. Here, we retrap $N=0$ or $N=1$ molecules in their lowest
hyperfine state with two beams with $\beta=0^\circ$ and total intensity
$I=36$~kW~cm$^{-2}$. The intensity of one of the beams is then modulated
sinusoidally by $\pm20\%$ for 1\,s. When the modulation frequency is twice the
trapping frequency, we resonantly heat the molecules and observe evaporative
loss from the trap as shown in Fig.~\ref{fig:Isotropic}(c). If the AC Stark
shift is linear, the trap frequency $\omega$ is proportional to
$\sqrt{\alpha/m}$, where $m$ is the mass. We compare the trap frequency for
molecules with those for $^{87}$Rb and $^{133}$Cs atoms in a dipole trap of the
same intensity (dashed lines in Fig.~\ref{fig:Isotropic}(c)) to find the
absolute polarizabilities of the molecules in both states. The value for $N=1$
is corrected for the contribution of $\alpha^{(2)}$ obtained above for
$\beta=0^{\circ}$. The two parametric heating results agree with one another
and with the spectroscopic result within experimental uncertainty. We find an
uncertainty-weighted average value
$\alpha^{(0)}/4\pi\epsilon_0=8.8(1)\times10^{2}~a_{0}^{3}$, in reasonable
agreement with theoretical predictions~\cite{Kotochigova:2006,
Kotochigova:2010, Vexiau:2012}. Note that the parametric heating approach does
not require knowledge of the absolute intensity of the trapping beams and thus
gives a smaller uncertainty in $\alpha^{(0)}$.

\begin{figure}[tb]
\centering
\includegraphics[width=\linewidth]{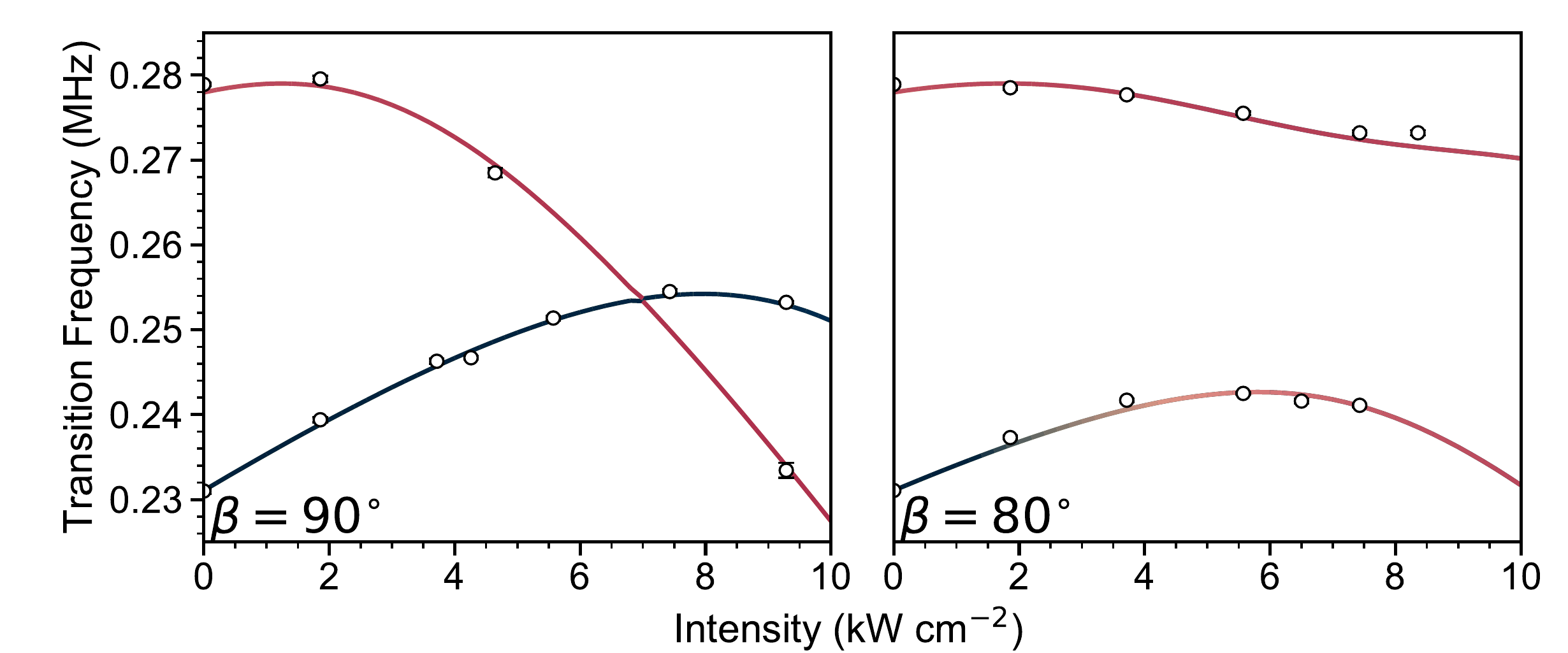}
\caption{\label{fig:AvoidedCrossing} Transition frequencies from $N=0, M_F=5$
to the two lowest energy hyperfine states of $N=1$ for $\beta=90^{\circ}$ and
$\sim80^{\circ}$. Transition strengths are coded as in Figs.~\ref{fig:Beta90}
and~\ref{fig:Beta00}.}
\end{figure}

The avoided crossings between laser-dressed levels can cause trap loss if the
molecules undergo Landau-Zener transitions to different hyperfine states as
they move around the trap, or if the intensity of the dipole trap is changed
dynamically. This is particularly important when retrapping molecules that have
been manipulated in free space. Since in our experiment trapping requires a
minimum laser intensity around 15~kW~cm$^{-2}$, such losses occur for example
when attempting to retrap the spin-stretched $N=1, M_F=6$ state with
$\beta=90^\circ$. Avoided crossings will also produce anharmonic and
anisotropic trapping potentials, which may result in complicated density
profiles for molecular clouds and cause coupling between translational and
rotational degrees of freedom in optical lattices. The strengths of avoided
crossings may be tuned by varying $\beta$; for example,
Fig.~\ref{fig:AvoidedCrossing} shows how the avoided crossing between the two
lowest-energy hyperfine levels of $N=1$ varies due to a change in
laser polarization of approximately 10$^{\circ}$. Understanding the avoided crossings will allow us to
identify optimum laser intensities and polarizations for optical trapping.
Furthermore, given sufficient broadening of the avoided crossing and precise
control of the laser intensity, it may be possible to traverse the avoided
crossings in a controlled manner during the retrapping of molecules. This may
allow access to hyperfine states that are not easily produced with microwave
transfer.

A further consequence of the AC Stark effect is that the frequencies of
microwave transitions depend on the position within an optical trap. This has
important ramifications for the design of experiments to achieve coherent
control of trapped molecules. Neyenhuis \emph{et al.}\ demonstrated coherence
times in Ramsey interferometry up to 1.5\,ms in $^{40}$K$^{87}$Rb by optimizing
a ``magic angle" between the magnetic field and the laser
polarization~\cite{Neyenhuis:2012}. To achieve longer coherence times, it is
desirable to find excited states that are parallel to the ground state as a
function of laser intensity. This condition is met at the turning point of an
avoided crossing.  This will make it possible to achieve longer coherence times
by controlling laser intensity as well as polarization.


In summary, we have completely characterized the anisotropic polarizability of
$^{87}$Rb$^{133}$Cs for $\lambda=1550$~nm. We have measured microwave spectra
of several hyperfine components of the the $N=0\rightarrow 1$ microwave
transition as a function of laser intensity and used them to extract precise
values of the anisotropic component $\alpha^{(2)}$ of the molecular
polarizability. We have supplemented this with parametric heating and
spectroscopic measurements to determine the isotropic component $\alpha^{(0)}$.
We have discovered a subtle interplay between the AC Stark effect and the
hyperfine structure, which produces a rich and complex pattern of avoided
crossings between levels as a function of laser intensity and polarization.
Understanding this pattern has allowed us to trap molecules in well-defined
hyperfine states and control their polarizability. This lays the foundations
for enhanced coherent microwave control of the internal state of polar
molecules confined in optical traps and lattices, which will underpin many
exciting proposed applications of ultracold molecules.

\begin{acknowledgments}
This work was supported by the U.K. Engineering and Physical Sciences Research
Council (EPSRC) Grants No.\ EP/H003363/1, EP/I012044/1, EP/P008275/1 and EP/P01058X/1. JA
acknowledges funding by the Spanish Ministry of Science and Innovation Grants
No.\ CTQ2012-37404-C02, CTQ2015-65033-P, and Consolider Ingenio 2010
CSD2009-00038. The experimental results and analysis presented in this paper
are available at DOI:10.15128/r13x816m612.
\end{acknowledgments}


\bibliography{RbCsReferences}

\end{document}